\documentclass[12pt,a4paper]{iopart}

\usepackage{graphicx}
\usepackage{color}
\usepackage{mathrsfs} 
\usepackage{dsfont}
\usepackage{iopams}  

\usepackage[bookmarks=true,
            bookmarksopen=true,colorlinks=false]{hyperref}

\newcommand{\eqref}{\eref}
\newcommand{\dd}{\mathrm{d}}

\newcommand{\R}{\mathds R}

\newcommand{\eps}{\varepsilon}

\newcommand{\Scri}{\mathscr{I}}

\begin{document}

\title[Fully pseudospectral solution of the conformally invariant wave equation II.]{Fully pseudospectral solution of the conformally invariant wave equation near the cylinder at spacelike infinity. II: Schwarzschild background}

\author{J\"org Frauendiener$^{1,2}$ and J\"org Hennig$^1$}
\address{$^1$Department of Mathematics and Statistics,
           University of Otago,
           PO Box 56, Dunedin 9054, New Zealand}
\address{$^2$Current address: Department of Mathematics, University of Oslo, Norway}
\eads{\mailto{joergf@maths.otago.ac.nz} and \mailto{jhennig@maths.otago.ac.nz}}

\begin{abstract}
It has recently been demonstrated (Class.\ Quantum Grav.\ {\bf 31}, 085010, 2014) that the conformally invariant wave equation on a Minkowski background can be solved with a fully pseudospectral numerical method. In particular, it is possible to include spacelike infinity into the numerical domain, which is appropriately represented as a cylinder, and highly accurate numerical solutions can be obtained with a moderate number of gridpoints. In this paper, we generalise these considerations to the spherically-symmetric wave equation on a Schwarzschild background. In the Minkowski case, a logarithmic singularity at the future boundary is present at leading order, which can easily be removed to obtain completely regular solutions. An important new feature of the Schwarzschild background is that the corresponding solutions develop  logarithmic singularities at infinitely many orders. 
This behaviour seems to be characteristic for massive space-times. In this sense this work is indicative of properties of the solutions of the Einstein equations near spatial infinity.  The use of fully pseudospectral methods allows us to still obtain very accurate numerical solutions, and the convergence properties of the spectral approximations reveal details about the singular nature of the solutions on spacelike and null infinity. These results seem to be impossible to achieve with other current numerical methods. Moreover, we describe how to impose conditions on the asymptotic behaviour of initial data so that the leading-order logarithmic terms are avoided, which further improves the numerical accuracy.\\[2ex]{}
{\it Keywords\/}: asymptotic structure, numerical relativity, collocation methods\\[-9.5ex]
\end{abstract}

\pacs{04.20.Ha, 04.25.D-, 02.70.Jn}

\section{Introduction\label{sec:intro}}

The study of 
the role of conformal structures in the large scale behaviour of solutions to the Einstein equations has a long tradition in general relativity, and was initiated by Roger Penrose's pioneering work in the 1960s \cite{Penrose1964,Penrose1965}. 
Of course, conformal transformations were already applied in differential geometry much earlier, probably for the first time by Hermann Weyl in 1918 \cite{Weyl1918}.
Nowadays, it is well known that conformal infinity is crucial for a rigorous description of gravitational waves. Most importantly, the Einstein equations themselves can be reformulated in terms of quantities that characterise the conformal structure and are regular even at ``infinity'', which leads to Helmut Friedrich's conformal field equations \cite{Friedrich1983,Friedrich1986}. Yet another, further refined reformulation of the field equations, which avoids certain mathematical difficulties at spacelike infinity $i^0$, is again due to Friedrich: the generalised conformal field equations \cite{Friedrich1998}. An important ingredient is a novel representation of $i^0$ as a cylinder $I$ that connects past and future null infinity $\Scri^\pm$. In this way, the behaviour of fields near spacelike infinity can be appropriately resolved. For comprehensive overviews of conformal methods and detailed references, we refer the interested reader to \cite{Frauendiener2004} and \cite{Kroon2016}.

In order to understand basic properties of fields near $i^0$, and, in particular, to study how they can be reconstructed numerically, it is most appropriate to adopt the formulation in which $i^0$ is blown up to the above-mentioned cylinder. However, instead of solving the rather complicated full system of the (generalised) conformal field equations, we consider the conformally invariant wave equation 
\begin{equation}\label{eq:CWE}
 g^{ab}\nabla_a\nabla_b f-\frac{R}{6}f=0
\end{equation}
for a scalar function $f$ as a toy model, where $g$ is a fixed metric and $R$ the corresponding scalar curvature. While this equation is much simpler, it has a similar structure of characteristics, and it is assumed to already mirror important features and potential numerical difficulties of the full problem. In the case of a Minkowski background, and under the assumption of spherical symmetry, this problem was successfully solved in \cite{FrauendienerHennig2014} with a fully pseudospectral scheme, which means that spectral expansions are used with respect to space \emph{and} time. The main result was that highly accurate solutions close to machine accuracy can be obtained and that one observes spectral convergence, i.e.\ the error decays exponentially with the numerical resolution. 
 
In this paper, as the next logical step, we extend these results to a less trivial spacetime: we consider the spherically-symmetric wave equation on a Schwarzschild background. Again our main interest is in an appropriate description of the behaviour near spacelike infinity. This implies, in particular, that we first have to identify suitable coordinates that are sufficiently well-behaved near the cylinder $I$. 
In the Minkowski case, initial data subject to a regularity condition (namely \emph{asymptotically conformally static data} for which the initial time derivative vanishes at the cylinder) have a regular time development, since this condition achieves that a logarithmic singularity that could appear at highest order is removed. In contrast to that we will observe that certain logarithmic singularities (at infinitely many orders) are generally present for a Schwarzschild background. Hence it is important to study what impact this has on our numerical considerations.

We will solve the Schwarzschild wave equation with the same fully pseudospectral method that we also used in the Minkowski case. This scheme is based on Marcus Ansorg's approach to solving elliptic problems with pseudospectral methods \cite{Ansorg2002}, and it was first generalised to time-evolution problems in \cite{HennigAnsorg2009}. For a number of further applications of this method, we refer the reader to \cite{AnsorgHennig2011, Hennig2013, KalischAnsorg, MacedoAnsorg2014}. We will not go into numerical details in this paper, since the method is extensively discussed in the previous references. However, we briefly summarise the main idea of the method, which consists in the following steps:
\begin{enumerate}
 \item We map the physical domain to one (or several) unit square(s) by introducing spectral coordinates $(\sigma,\,\tau)\in[0,1]\times[0,1]$ such that the surface on which initial data are given corresponds to $\tau=0$. For the present application, one unit square turns out to be sufficient.
 \item We enforce the initial conditions by expressing the unknown function $f(\sigma,\tau)$ in terms of another unknown $f_2(\sigma,\tau)$ via
 $f(\sigma,\tau)= f_0(\sigma)+\tau f_1(\sigma)+\tau^2 f_2(\sigma,\tau)$.
 \item We choose spectral resolutions $n_\sigma$ and $n_\tau$ in spatial and temporal directions and approximate the new unknown $f_2$ in terms of Chebyshev polynomials $T_i$ in the form
 \begin{equation}
 f_2(\sigma,\tau)\approx\sum\limits_{i=0}^{n_\sigma-1}\sum\limits_{j=0}^{n_\tau-1}
        c_{ij}T_i(2\sigma-1)T_j(2\tau-1).
 \end{equation}
 \item We obtain an algebraic system of equations by requiring that the relevant differential equations (and boundary or regularity conditions, where applicable) are satisfied at a set of collocation points. For our purposes, a suitable choice are Gauss-Lobatto nodes
 $(\sigma_i,\tau_j)$, $i=0,\dots,n_\sigma-1$, $j=0,\dots,n_\tau-1$, defined by
 \begin{equation}
 \sigma_i=\sin^2\left(\frac{i\pi}{2(n_\sigma-1)}\right),\quad
  \tau_j=\sin^2\left(\frac{j\pi}{2(n_\tau-1)}\right).
 \end{equation}
 These have the advantage that gridpoints lie at all four boundaries $\sigma,\tau=0,1$.
 \item Starting from some initial guess, we solve this system iteratively with the Newton-Raphson method. In the present case of a \emph{linear} wave equation, we can simply choose the trivial solution $f_2\equiv 0$. For nonlinear equations, however, one usually needs to provide an initial guess sufficiently close to the solution, in order to guarantee convergence.
\end{enumerate}

This paper is organised as follows. In Sec.~\ref{sec:nearinf}, we introduce coordinates that are suitable for a discussion of the wave equation near spacelike infinity. In these coordinates, we first reconstruct a simple test solution numerically, which is regular everywhere. 
Then we show that general solutions suffer from logarithmic singularities at $I^+$, where the sets $I$ and $\Scri^+$ approach each other.
We also show that the development of initial data that are chosen subject to certain regularity conditions avoids the leading-order singularities, and we study how the logarithmic terms at different orders influence the numerical accuracy. 
Afterwards, in Sec.~\ref{sec:horizon}, we include the event horizon into the numerical domain. To this end, we show how a coordinate singularity of the previous coordinates, which prevents access to the horizon, can be removed, and we numerically solve the wave equation in the modified coordinates. Finally, in Sec.~\ref{sec:discussion}, we summarise and discuss our results.

\section{Numerical studies near spacelike infinity}\label{sec:nearinf}
\subsection{Coordinates and the conformally invariant wave equation\label{sec:coords}}

In order to investigate and numerically solve the conformally invariant wave equation near spacelike infinity, we first need to choose suitable coordinates that are sufficiently well-behaved in the sense that they allow us to construct accurate numerical solutions. Ideally, one could envision coordinates that cover the entire exterior of the black hole --- and probably even further regions of the maximally extended Schwarzschild solution. The well-known standard compactification and its variations simply shrink the Kruskal-Szekeres coordinates by mapping the corresponding null coordinates to a finite domain, see, e.g., \cite{FrolovNovikov,GriffithsPodolsky,MTW}. The resulting metrics, however, are typically not conformally related to another metric with desirable geometric properties, in particular near infinity, and at least some degree of regularity of the metric will be lost at the conformal boundary. A promising new family of coordinate systems that achieves an analytic conformal compactification has been constructed in \cite{HalacekLedvinka} with particular view on suitability for numerical computations. While $i^0$ is still treated as a point in these coordinates, it is possible to blow it up to the cylinder $I$ with an additional coordinate transformation. Unfortunately, even though these coordinates behave better near null infinity (as shown in a direct comparison with standard compactifications in \cite{HalacekLedvinka}), they turn out not to be optimal for our treatment of spacelike infinity. The problem is that the metric contains terms that behave like $\eps^2\ln\eps$ near the cylinder, which is assumed to be located at $\eps=0$. This is only a weak logarithmic singularity with bounded function values and derivatives, but second derivatives blow up as $\eps\to0$. Since spectral methods are very sensitive to this type of irregularity, these coordinates are not a good choice for our purposes. 
As a consequence, we have to be satisfied with choosing other coordinates that cover only a certain portion of the exterior vacuum region, but are suitable for our numerical computations. However, since our main interest lies in the behaviour near the cylinder, this appears to be an acceptable restriction.

\begin{figure}\centering
 \includegraphics[width=12cm]{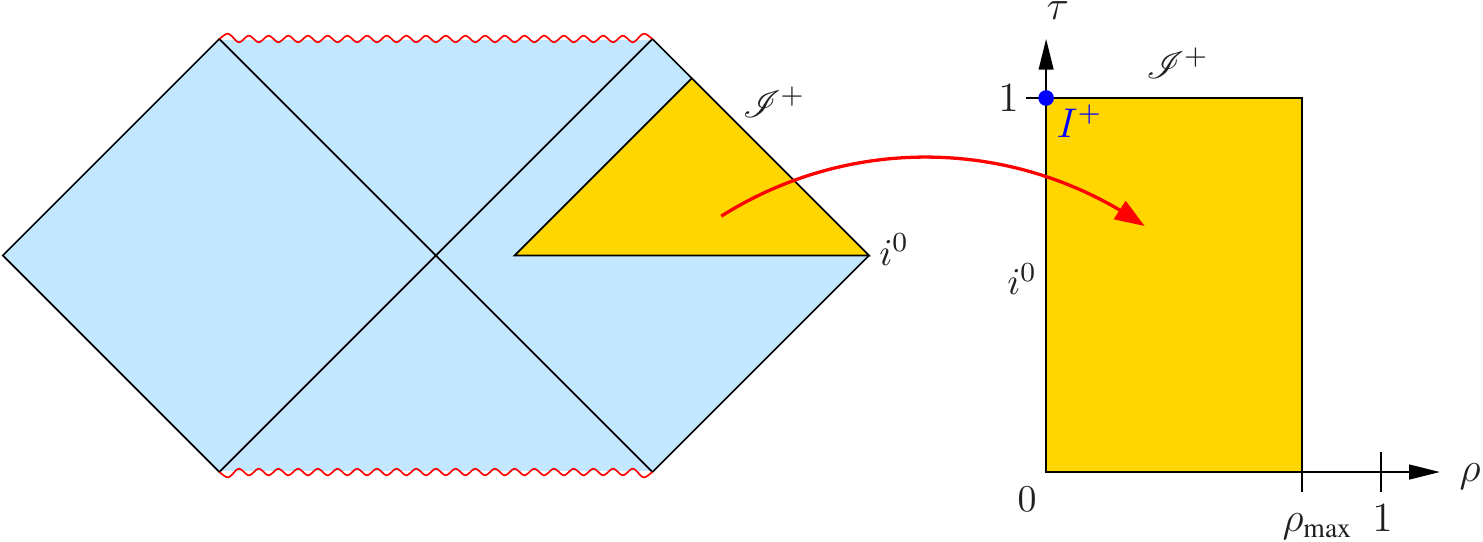}
 \caption{Illustration of the domain covered by the coordinates $\rho$ and $\tau$.
 \label{fig:ConfDiag}}
\end{figure}

We first consider coordinates that were introduced by Friedrich \cite{Friedrich2004}. These are adapted to one family of radial null geodesics (instead of both families, as in the Kruskal-Szekeres coordinates), and a conformal metric is chosen according to the conformal cn-gauge\footnote{The \emph{cn-gauge} reduces the degree of freedom of choosing a conformal spacetime by imposing that orthonormal frames are transported in a certain way along conformal geodesics, see \cite{Friedrich1998,Friedrich2004} for details.}. Starting from the Schwarzschild metric $\tilde g$ in isotropic coordinates $(\tilde t,\tilde r,\theta,\phi)$,
\begin{equation}
 \tilde g=\left(\frac{1-\frac{m}{2\tilde r}}{1+\frac{m}{2\tilde r}}\right)^2
 \dd {\tilde t}^{\,2}-\left(1+\frac{m}{2\tilde r}\right)^4(\dd\tilde r^2+\tilde r^2\dd\sigma^2)
\end{equation}
with $\dd\sigma^2:=\dd\theta^2+\sin^2\theta\,\dd\phi^2$, we compactify the radial coordinate $\tilde r$ and rescale the time coordinate $\tilde t$ by setting
\begin{equation}\label{eq:trans1}
 r=\frac{m}{2\tilde r},\quad t=\frac{2\tilde t}{m}.
\end{equation}
With the additional transformation 
\begin{equation}\label{eq:trans2}
 t=\int_r^\rho\frac{\dd s}{F(s)},\quad r=\rho(1-\tau),
  \quad\textrm{where}\quad
   F(s)=\frac{s^2(1-s)}{(1+s)^3},
\end{equation}
we finally arrive at coordinates $(\tau,\rho,\theta,\phi)$ such that $\rho$ is a null coordinate and in which spacelike infinity is blown up. Note that the integral in the above coordinate transformation can be evaluated to give
\begin{equation}\label{eq:texplicit}
 t=-\rho\tau+\frac{\tau}{r}-4\ln(1-\tau)-8\ln\frac{1-\rho}{1-r}, 
\end{equation}
but this explicit form is not needed in the following. The metric can now be written as $\tilde g=\Theta^{-2}g$ in terms of the conformal metric
\begin{equation}\label{eq:conmet}
 g=\frac{2}{\rho}A\,\dd\rho\,\dd\tau-\frac{1-\tau}{\rho^2}A[2-(1-\tau)A]\dd\rho^2-\dd\sigma^2,
\end{equation}
where
\begin{equation}\label{eq:defA}
 A:=\frac{F(r)}{(1-\tau)^2F(\rho)}\equiv \frac{(1-r)(1+\rho)^3}{(1-\rho)(1+r)^3},
\end{equation}
and with the conformal factor
\begin{equation}\label{eq:Theta}
 \Theta=\frac{2r}{m(1+r)^2}.
\end{equation}
We consider the new coordinates in the intervals $0<\rho\le\rho_\mathrm{max}<1$ and $0\le\tau<1$. The restriction of $\rho$ to values below a maximum $\rho_\mathrm{max}$ is necessary, because the transformation \eqref{eq:trans2} introduces a coordinate singularity at $\rho=1$, due to $F(1)=0$. This is also clear from the explicit relation \eqref{eq:texplicit}. As a consequence, the conformal metric \eqref{eq:conmet} is singular at $\rho=1$, where $A$ diverges, see \eqref{eq:defA}. Since the event horizon of the Schwarzschild black hole is located at $\tilde r=m/2$, corresponding to $r=1$, i.e.\ $\rho=1/(1-\tau)\ge 1$, the region that can be described with these coordinates does not extend to the horizon. Instead, the coordinates cover a domain as indicated in the standard Penrose diagram in Fig.~\ref{fig:ConfDiag}. 
The boundaries of our $\rho$-$\tau$ domain have the following correspondence to regions of the extended Schwarzschild spacetime:
\begin{enumerate}
 \item $\tau=0$, $0\le\rho\le\rho_\mathrm{max}$:
 The initial slice on which we prescribe the data $f$ and $f_{,\tau}$.
 \item $\rho=\rho_\mathrm{max}$, $0\le\tau\le 1$:
 This boundary, like all curves $\rho=\mathrm{constant}$, is an outgoing null geodesic and a characteristic of the corresponding wave equation. Hence no information can enter our domain through this boundary, so that no boundary conditions are required there. This is particularly convenient for our numerical computations as we avoid to impose ``artificial boundary conditions'', as is done so often in numerical simulations, even though this is physically rather inappropriate, since usually no information about the unknown functions is available at such boundaries that would justify to impose any particular conditions there. The exact location of the left boundary of the triangular region in Fig.~\ref{fig:ConfDiag} depends on the choice of $\rho_\mathrm{max}$.
 \item $\tau=1$, $0<\rho\le\rho_\mathrm{max}$: (A part of) Schwarzschild future null infinity $\Scri^+$.
 \item $\rho=0$, $0\le\tau<1$: The (future half of) the cylinder representation of spacelike infinity.
 \item $\rho=0$, $\tau=1$: The ``critical set'' $I^+$.
\end{enumerate}

We are now in a position to formulate the conformally invariant wave equation in terms of the $\rho$-$\tau$ coordinates. From a physical point of view, the equation is initially valid for finite values of the original Schwarzschild coordinates $\tilde t$ and $\tilde r$. In terms of our compactified coordinates, we extend this domain by imposing the same equation also at the conformal boundaries $\rho=0$ and $\tau=1$. Note that, while standard hyperbolic PDE theory guaranties existence of unique solutions in the original, ``physical'' set $0\le\tau<1$, $0<\rho\le\rho_\mathrm{max}$, existence on the extended set is not immediately clear (due to the particular degeneracy of the equation at the boundaries, in particular, at $I$). This gap, however, is closed with our numerical investigations, which provide (numerical approximations to) solutions with a certain degree of smoothness on the extended domain.

In order to formulate the wave equation \eqref{eq:CWE}, we first compute the Ricci scalar in these coordinates, which turns out to be 
$R=-24r/(1+r)^2 \equiv -12m\Theta$. With this we obtain the wave equation
for a function $f=f(\tau,\rho)$,
\begin{equation}\fl\label{eq:CWE1}
 (1-\tau)[2-(1-\tau)A]\,f_{,\tau\tau} +2\rho\, f_{,\tau\rho}
 -2\left[1-\frac{1-2r}{1-r^2}(1-\tau)A\right]f_{,\tau}+\frac{4r A}{(1+r)^2}\,f=0.
\end{equation}
The metric $g$ in \eqref{eq:conmet} is real analytic when $\rho\ne0$. This property guarantees that \eqref{eq:CWE1} is hyperbolic on the physical part of the computational domain. The theory of hyperbolic equations then implies that the initial value problem based on the hypersurface $\tau=\mathrm{constant}$ is well posed for initial data on $0<\rho<\rho_\mathrm{max}$ which are sufficiently regular.

The equation also makes sense for $\rho=0$ but it loses hyperbolicity there. Therefore, we need to discuss this location separately. The discussion is significantly simplified by the fact that $\rho=0$ is a total characteristic of the equation. This can be seen as follows.

If we consider \eqref{eq:CWE1} in the limit $\rho\to0$, i.e.\ at the cylinder, then it reduces to the intrinsic equation $(1-\tau^2)f_{,\tau\tau}-2\tau f_{,\tau}=0$. This immediately implies that $f_{,\tau}(\tau,0)=c/(1-\tau^2)$ for some integration constant $c$. Hence the solution $f$ generally diverges at $I^+\ (\rho=0,\ \tau=1$), where $I$ and $\Scri^+$ approach each other, unless we choose initial data with 
\begin{equation}\label{eq:incon}
 f_{,\tau}(\tau=0, \rho=0)=0, 
\end{equation}
for which $c=0$ follows. This is exactly the same situation as for the conformally invariant wave equation on a Minkowski background \cite{FrauendienerHennig2014}. As in the Minkowski case, we will exclusively consider initial data subject to this regularity condition. Note, however, that we will later observe an interesting difference to the situation with a Minkowski background: while initial data subject to this regularity condition guarantee a regular solution in the Minkowski case, the solutions to the Schwarzschild wave equation generally still contain higher-order logarithmic singularities at $I^+$. This will be revealed by a more detailed study of the behaviour of $f$ near the cylinder in Sec.~\ref{sec:expansion} below.

For our choice of initial data subject to \eqref{eq:incon}, the solution $f$ satisfies $f_{,\tau}=0$ on the entire cylinder $\rho=0$, 
and we will impose this equation there.\footnote{Note that the cylinder $I$ is a characteristic surface. Consequently, from the analytical point of view, boundary conditions are not required at $I$ and we could impose the wave equation itself. However, since the wave equation reduces to an intrinsic equation there, which, for our choice of initial data, is equivalent to $f_{,\tau}=0$, we can still choose to impose this condition as a numerical boundary condition}.
For an interpretation of this condition, we can reformulate it in terms of the original Schwarzschild coordinates $\tilde t$ and $\tilde r$, with respect to which it becomes $f_{,\tilde t}\to0$ as $\tilde r\to\infty$ for finite $\tilde t$. If $f$ has an expansion $f(\tilde t,\tilde r)=\psi_0(\tilde t)+\psi_1(\tilde t)r^{-1}+\psi_2(\tilde t)r^{-2}+\dots$, then this condition implies $\psi_0=\mathrm{constant}$, i.e.\ it enforces that $f$ has a unique limit at spacelike infinity irrespective of the spacelike slice along which it is approached.

Note that, at $\tau=\rho=0$, the condition $f_{,\tau}=0$ is identically satisfied for our initial data. Instead, we can impose $f_{,\tau\tau}=0$ at this particular point, which immediately follows from the other condition by differentiation.

Furthermore, exactly as in the Minkowski case \cite{FrauendienerHennig2014}, it turns out that numerical stability requires to also choose another condition than $f_{,\tau}=0$  at the special point $\rho=0$, $\tau=1$, i.e.\ at $I^+$ (provided that we decide to include $\Scri^+$ into the numerical domain). 
The reason why an extra condition is required is the following. At $\tau=1$, the wave equation implies the intrinsic equation $\rho f_{,\tau\rho}-f_{,\tau}=0$ for sufficiently smooth solutions. It follows that $f_{,\tau}=c\rho$ there for some constant $c$. Due to the intrinsic nature of the equation at $\tau=1$, the problem at $\tau=1$ is decoupled from the problem at $\tau<1$, so that this constant is completely undetermined from the numerical point of view. Hence the pseudospectral method cannot converge to a well-defined solution. However, if we provide an alternative condition that fixes $c$, then the numerical method will converge.
The required extra condition, of course, must be a consequence of the wave equation --- certainly we cannot choose just any condition. Similarly to the Minkowski situation, the extra condition is obtained by differentiating the wave equation with respect to the radial coordinate $\rho$. At $\rho=0$, this gives $(1+\tau)f_{,\tau\tau\rho}+2f_{,\tau\rho}+4f=0$. Since $f(\tau,0)=:f_0=\mathrm{constant}$ on the cylinder, this is an intrinsic equation for the mixed derivative $g(\tau):=f_{,\tau\rho}(\tau,0)$. The solution can be expressed in terms of the values $f_0$ and $g_0:=g(0)\equiv f_{,\tau\rho}(0,0)$, 
\begin{equation}
 g(\tau)=\frac{g_0+2f_0}{(1+\tau)^2}-2f_0.
\end{equation}
Hence we know the values of $f_{,\tau\rho}$ everywhere on the cylinder. At the particular point $I^+$, we can prescribe the correct value for this mixed derivative (which is identical with the above-mentioned constant $c$, which is therefore fixed), i.e.\ we can use the boundary condition $f_{,\tau\rho}(1,0)=(g_0-6f_0)/4$, where $f_0$ and $g_0$ can be read off from the initial data. In summary, at $\rho=0$, i.e.\ at the cylinder, we impose the following boundary conditions:
\begin{equation}
 \textrm{at }\rho=0:\quad
 \cases{
  f_{,\tau\tau}=0, & $\tau=0$,\\
  f_{,\tau\rho}=\frac14(g_0-6f_0), & $\tau=1$,\\
  f_{,\tau}=0, & otherwise.}
\end{equation}
Above we have mentioned how the condition $f_{,\tau}=0$ translates into the Schwarzschild coordinates $\tilde t$, $\tilde r$. A similar interpretation of the extra condition at $\tau=1$ would be interesting, but may be difficult to obtain, as this would require to consider the simultaneous limit $\tilde r\to\infty$, $\tilde t\to\infty$, where the limit is performed in exactly such a way that the special point $I^+$ is approached.

Finally, another boundary of our numerical domain is $\rho=\rho_\mathrm{max}$. As mentioned above, no boundary condition is required there as this is a characteristic of the wave equation. Instead, we just impose the wave equation itself.

\subsection{A simple test solution}
In order to find out whether the fully pseudospectral scheme can be applied to the Schwarzschild wave equation, we first try to numerically reproduce a simple exact solution. This can be obtained by looking for time-independent solutions with respect to the original isotropic Schwarzschild coordinates $(\tilde t,\tilde r)$. In these coordinates, the conformally invariant wave equation for a function $\tilde f=\tilde f(\tilde r)$ reads
\begin{equation}
 \left[\left(\tilde r^2-\frac{m^2}{4}\right)\tilde f_{,\tilde r}\right]_{,\tilde r}=0.
\end{equation}
This equation can be readily integrated to obtain
\begin{equation}
 \tilde f(\tilde r)=c\ln\frac{2\tilde r-m}{2\tilde r+m}+d
\end{equation}
with two integration constants $c$ and $d$. The corresponding solution $f$ in our compactified coordinates $(\tau,\rho)$ is obtained from $\tilde f$ via $f=\Theta^{-1}\tilde f$, since (in four spacetime dimensions) the conformally invariant wave equation has the conformal weight $-1$. Hence we get
\begin{equation}
 f(\tau,\rho)=\frac{m(1+r)^2}{2r}\left(c\ln\frac{1-r}{1+r}+d\right),
 \quad r=\rho(1-\tau).
\end{equation}
In this form, the solution has a nontrivial dependence on the compactified time coordinate $\tau$. 

For regularity at the cylinder we have to choose $d=0$. Moreover, without loss of generality (given that we study a linear equation), we can choose $c=2/m$ to finally obtain
\begin{equation}\label{eq:testsol}
 f(\tau,\rho)=\frac{(1+r)^2}{r}\ln\frac{1-r}{1+r}
\end{equation}
as an exact solution that is regular in the entire domain $0\le\rho\le\rho_\mathrm{max}$, $0\le\tau\le 1$.

\begin{figure}\centering
 \includegraphics[width=8.5cm]{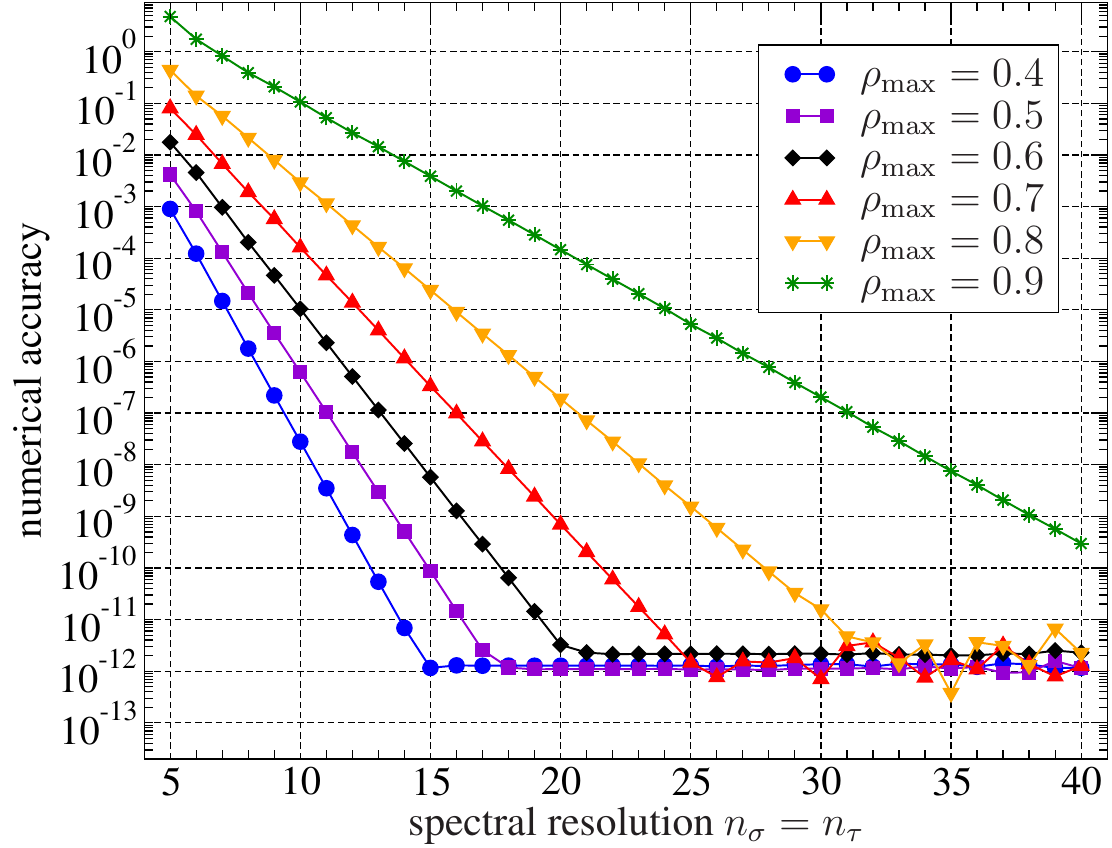}
 \caption{Convergence plot for the numerical reconstruction of the test solution \eqref{eq:testsol}. The numerical error is shown as a function of the spectral resolutions, where the same resolutions in spatial and time directions have been chosen, $n_\sigma=n_\tau$. \label{fig:TestSol}}
\end{figure}

Now we try to reconstruct this solution from the corresponding initial data $f(0,\rho)$ and $f_{,\tau}(0,\rho)$, which can be read off from \eqref{eq:testsol}. In particular, we are interested in the numerical error, which we compute as the largest absolute difference between the numerical approximation and the exact solution, $\max|f_\mathrm{numerical}-f_\mathrm{exact}|$. Here, the maximum is approximated by comparing the solutions at $100\times100$ equidistant points, where the numerical values are obtained via Chebyshev interpolation. The results are displayed in Fig.~\ref{fig:TestSol}, 
where the error is shown for different spectral resolutions and for different choices of the coordinate range specified by the value of $\rho_\mathrm{max}$. We observe that, for each choice of $\rho_\mathrm{max}$, the error as a function of the resolution approximately follows a straight line until saturation is reached close to machine accuracy (about 16 figures for our double-precision code). A straight line in a logarithmic plot corresponds to an exponentially decaying error, i.e.\ to spectral convergence. This is exactly what is typically expected for (pseudo)spectral methods, provided the solution is sufficiently well-behaved. If we compare the curves that are obtained for different values of $\rho_\mathrm{max}$, then clearly higher resolutions are required to achieve the same error for a larger $\rho_\mathrm{max}$. Nevertheless, for sufficient resolutions, the errors reach the same order of magnitude of about $10^{-11}$ to $10^{-12}$, corresponding to about $11$ to $12$ significant figures in the numerical solutions. It was to be expected that a larger value of $\rho_\mathrm{max}$ requires a higher resolution, since the function $f$ then needs to be represented on a larger domain. However, an even more important reason for the requirement of higher resolutions is that larger $\rho_\mathrm{max}$ bring the numerical domain closer to the singularity of the term $\ln(1-r)$ appearing in the exact solution \eqref{eq:testsol}. Since the solution has steep gradients near the boundary $\rho=\rho_\mathrm{max}$, more Chebyshev polynomials are naturally required to represent the function accurately. Nevertheless, we can still reach an accuracy close to machine accuracy for just a moderate number of collocation points, even if the domain is relatively close to the coordinate singularity at $\rho=1$. Note that we could even further improve the numerical performance by appropriately stretching our coordinates in a vicinity of $\rho=1$, which effectively places more collocation points near the boundary. This is described in detail in \cite{MeinelBook}. For our purposes, however, this is not really necessary. Firstly, we are mainly interested in a vicinity of spacelike infinity, so there is no need to choose $\rho_\mathrm{max}$ very close to $1$. Secondly, the singularity at $\rho=1$ is a coordinate singularity without any physical meaning. Later we will modify our coordinates in order to remove that singularity, and hence there is no reason to study the wave equation close to a singularity that can be removed.

In summary, we find that we can solve the conformally invariant wave equations with the fully pseudospectral scheme. We observe spectral convergence and obtain highly accurate solutions close to machine accuracy. This is possible since our test solution \eqref{eq:testsol} is well-behaved in the entire numerical domain. In the next subsection, however, we will see that general solutions usually develop logarithmic singularities at the point $I^+$. Indeed, our exact solution \eqref{eq:testsol} and multiples of it may well be the only solutions without such singularities. Hence it is important to understand the general behaviour of solutions near $i^0$ and to test to what extent the numerical method works in those cases.

\subsection{General behaviour near spacelike infinity}\label{sec:expansion}

Next we consider the case of general initial data at $\tau=0$, which usually give rise to solutions that are less regular than our explicit example solution \eqref{eq:testsol}. 

In order to analyse the behaviour near the cylinder, we expand the solution $f$ in a Taylor series about $\rho=0$ with $\tau$-dependent coefficients,
\begin{equation}\label{eq:expan}
 f(\tau,\rho)=\phi_0(\tau)+\phi_1(\tau)\rho+\phi_2(\tau)\rho^2+\phi_3(\tau)\rho^4+\dots
\end{equation}
Plugging this form of $f$ into the wave equation \eqref{eq:CWE1}, we obtain a sequence of ordinary differential equations for the functions $\phi_0$, $\phi_1$, $\dots$ The first four of these will be discussed and solved in the following. If we abbreviate the left hand side of \eqref{eq:CWE1} with $W$, so that the wave equation can be written as $W=0$, we obtain at the cylinder, i.e.\ in the limit $\rho\to0$,
\begin{equation}
 \lim_{\rho\to0} W=(1-\tau^2)\ddot\phi_0-2\tau\dot\phi_0=0,
\end{equation}
where a dot denotes differentiation with respect to $\tau$.
The general solution 
\begin{equation}
 \phi_0(\tau)=c_0+c_1\ln\frac{1-\tau}{1+\tau} 
\end{equation}
with integration constants $c_0$ and $c_1$ diverges at $\tau=1$, which, at $\rho=0$, corresponds to the point $I^+$. However, since $\dot\phi_0(0)=-2c_1$, we can achieve that the singular term vanishes, if we choose initial data with
\begin{equation}\label{eq:regcon0}
 f_{,\tau}(0,0) \equiv \dot\phi_0(0)=0
\end{equation}
so that $c_1=0$. This is exactly the same condition for the initial data that we discussed earlier [cf.\ \eqref{eq:incon}], and which is required for bounded function values at $I^+$. Hence we can eliminate the leading-order singularity at $I^+$ with an appropriate choice of initial data. Unlike the corresponding Minkowski case, where this was the only singular term at $I^+$, we will find weaker logarithmic singularities at higher orders in the Schwarzschild case.

Assuming that initial data with $c_1=0$ have been chosen, we next examine the equation for $\phi_1$. This is obtained from
\begin{equation}
 \lim_{\rho\to 0}\frac{W}{\rho} =(1-\tau^2)\ddot\phi_1+2(1-\tau)\dot\phi_1+4c_0(1-\tau)=0,
\end{equation}
which has the general solution
\begin{equation}\label{eq:phi1}
 \phi_1(\tau)=-2c_0\tau-\frac{2c_0+c_2}{1+\tau}+c_3.
\end{equation}
Since this function is regular for all $0\le\tau\le 1$, no further singularities are introduced at this order.\footnote{Note that \eqref{eq:phi1} is generally still singular at $\tau=-1$. If we want to consider the solution on a larger time interval that includes $\tau=-1$, then further restrictions need to be imposed on the initial data in order to remove singularities. For example, initial data with $(2f+f_{,\tau\rho})|_{\tau=\rho=0}\equiv 2c_0+c_2=0$ lead to a solution with regular $\phi_1$, since the $1/(1+\tau)$-term in \eqref{eq:phi1} then disappears. In the following, however, we will restrict ourselves to domains with $\tau\ge0$.}

Once $\phi_1$ is chosen as in \eqref{eq:phi1}, the next equation becomes
\begin{eqnarray}\fl\nonumber
 \lim_{\rho\to 0}\frac{W}{\rho^2}
 &=& (1-\tau^2)\ddot\phi_2+2(2-\tau)\dot\phi_2
  -8c_0\frac{(1-\tau)(\tau^4+3\tau^3+3\tau^2-\tau+2)}{(1+\tau)^3}\\
 \fl\
 && -8c_2\frac{(1-\tau)^2}{(1+\tau)^3}+4c_3(1-\tau)=0.
\end{eqnarray}
The general solution to this equation again contains a logarithmic term,
\begin{eqnarray}\fl\nonumber
 \phi_2(\tau) 
 &=& 4(2c_0-c_3)\left(\frac{1-\tau}{1+\tau}\right)^2\ln(1-\tau)
   +\frac{8(2c_0+c_2)\tau}{(1+\tau)^2}\ln(1+\tau)
   +\frac{c_4\tau}{(1+\tau)^2}+c_5\\
 \fl\nonumber
  &&  +\frac{2\left[c_0(2\tau^4+6\tau^3+6\tau^2+84\tau+36)+6c_2(1+2\tau)
    -3c_3(\tau^3+2\tau^2+10\tau+4)\right]}{3(1+\tau)^2}\\
 \fl
 &=& 4(2c_0-c_3)\left(\frac{1-\tau}{1+\tau}\right)^2\ln(1-\tau)
      +\textrm{regular terms}.
\end{eqnarray}
Here, the singular term is proportional to $(1-\tau)^2\ln(1-\tau)$, i.e.\ $\phi_2$ and $\dot\phi_2$ are bounded, but $\ddot\phi_2$ diverges as $\tau\to1$. This corresponds to a divergence of the fourth derivative $f_{,\tau\tau\rho\rho}$ at $I^+$. However, it is again possible to choose suitable initial data for which this term does not appear. Since
$(f_{,\rho}+f_{,\tau\rho})|_{\tau=\rho=0}=\phi_1(0)+\dot\phi_1(0)=c_3-2c_0$, 
we can achieve that the coefficient $2c_0-c_3$ of the singular term vanishes by choosing initial data subject to the additional regularity condition
\begin{equation}\label{eq:regcon2}
 (f_{,\rho}+f_{,\tau\rho})\Big|_{\tau=\rho=0}=0.
\end{equation}

Finally, we consider one further order, assuming that initial data are chosen such that singularities in all previous orders have been eliminated. The equation that results from $\lim_{\rho\to0}(W/\rho^3)=0$ is rather lengthy and we do not give it explicitly, but we observe that the general solution has the form
\begin{equation}
 \phi_3(\tau)=\frac83(6c_0+5c_2+3c_5)\left(\frac{1-\tau}{1+\tau}\right)^3\ln(1-\tau)+\textrm{regular terms}.
\end{equation}
This time we find a logarithmic term proportional to $(1-\tau)^3\ln(1-\tau)$, which corresponds to a singularity in the sixth derivative $f_{,\rho\rho\rho\tau\tau\tau}$.
Once again it is possible to choose initial data for which the coefficient of the singular term vanishes. Here this can be achieved with data subject to
\begin{equation}\label{eq:regcon3}
 (3f_{,\rho\rho}+14f_{,\rho}-36f)\Big|_{\tau=\rho=0}=0.
\end{equation}

These considerations illustrate that logarithmic terms are present in the solution near $I^+$, but they can be eliminated step-by-step by choosing initial data subject to more and more regularity conditions at $\tau=\rho=0$. 

For some general remarks on a systematic treatment of the singular behaviour of $\phi_n(\tau)$, we refer to \ref{sec:recursion}. In particular, we observe that, if all singularities up to the order $n-1$ have been eliminated, then the singular term at order $n$ is proportional to $(1-\tau)^n\ln(1-\tau)$. In other words, the trend observed here for the explicitly calculated $\phi$-functions continues in the same way at higher orders. The calculations in the appendix also show that the singularities are related to the non-vanishing mass $m$ of the Schwarzschild spacetime, and all singularities in $\phi_1$, $\phi_2$, $\dots$ disappear in the limit $m\to0$, leaving only a (possible) singularity in $\phi_0$ (depending on the initial data).

\subsection{Numerical studies in the general case}\label{sec:numgen}

How do we numerically solve the conformally invariant wave equation in situations where the logarithmic singularities from the previous subsection are present? In cases where the exact structure of a singularity is known, one can try to reformulate the problem entirely in terms of regular quantities. An example is the computation of rotating disks of dust, as discussed, e.g., in \cite{AnsorgMeinel2000} or \cite{HennigNeugebauerAnsorg}. The mass density of these disks behaves like a square root near the rim of the disk, so that it has unbounded derivatives. However, one can simply write the mass density as a square root times another function, where this other function turns out to be regular. Hence it is possible to reformulate the field equations using this regular function, and, consequently, one can again obtain highly accurate numerical solutions.

As a second example for removing singularities, we point to the numerical construction of five-dimensional black strings with pseudospectral methods in \cite{KalischAnsorg}. While the physical relevance of these configurations is rather questionable, they are certainly mathematically very interesting. It turns out that some of the unknown functions contain logarithmic terms of the form $\chi^l\ln^m\chi$, where $\chi$ is one of the coordinates and $l,m>0$ are constants. With a coordinate transformation $\chi=\chi_I\exp(1-1/\eta)$, $\chi_I=\mathrm{constant}$, these logarithmic terms  (and therefore the metric potentials) become $C^\infty$ (even though not analytic) functions of the new coordinate $\eta$, which is sufficient for highly accurate numerics.

In our case, unfortunately, both of the above regularisation procedures seem not to be applicable, as they require an exact knowledge of the structure of the singular terms. Since the calculations in the previous subsection only reveal the behaviour of $f$ and its  derivatives as $I^+$ is approached along the cylinder $\rho=0$, and not along any other curve, it is not clear if $f$ can be written as a combination of singular terms and one or more regular functions. Another difference between our situation and the above examples, where time-independent equilibrium configurations are studied, is that we have to solve a time-evolution problem. Even if $f$ could be expressed in terms of regular functions and well-defined singular terms, we would then also need to decompose the initial data into parts that could be used as initial data for those regular functions and parts that give rise to the singular terms. Since the initial data can be chosen as arbitrary, regular functions at $\tau=0$, it is not clear how the individual components could be extracted. (If the above hypothesis is true, according to which the only entirely regular solutions in our domain $0\le\tau\le 1$ and $0\le\rho\le\rho_\mathrm{max}$ may be the test solution \eqref{eq:testsol} and its multiples, then the ``regular portion'' of the initial data may correspond to initial data for this solution. But this was only a conjecture.)

Hence it appears to be impossible to eliminate the logarithmic terms from our solution via coordinate transformations or reformulations of the unknown function. Instead, we will try to solve the equation without reformulations and thereby test how the presence or absence of the leading-order singularities influences the numerical accuracy. However, we will only allow initial data subject to the regularity condition \eqref{eq:regcon0}, for which the highest-order singularity in $\phi_0$ is eliminated. This ensures that at least function values, even though not derivatives, are bounded.

In the following, we choose $\rho_\mathrm{max}=0.5$ and consider the family of initial data given by
\begin{equation}\label{eq:iniseq}
 \tau=0:\quad
 f=\frac{35+\beta}{18}+\sin(5\rho),\quad
 f_{,\tau}=(\alpha-5)\rho,
\end{equation}
which depends on two parameters $\alpha,\beta\in\R$. These parameters allow us to control which of the leading-order singularities will be present in the evolution of these data.
Since $f_{,\tau}(0,0)=0$ holds for all values of $\alpha$ and $\beta$, the leading singularity in $\phi_0$ is removed --- as required. We also obtain 
$(f_{,\rho}+f_{,\tau\rho})|_{\tau=\rho=0}=\alpha$. Hence the parameter $\alpha$ controls the next-order logarithmic singularity in $\phi_2$, cf.~\eqref{eq:regcon2}: for $\alpha=0$ there is no singularity in $\phi_2$, whereas $\phi_2$ diverges at $\tau=1$ for $\alpha\neq0$. In the latter case, the magnitude of $\alpha$ determines how large the coefficient of the logarithmic term is, i.e.\ how ``strong'' the singularity is. 
Finally, this family of initial data satisfies 
$(3f_{,\rho\rho}+14f_{,\rho}-36f)|_{\tau=\rho=0}=-2\beta$. Comparing with \eqref{eq:regcon3}, we see that we can eliminate the next-order logarithmic singularity by choosing $\beta=0$.

\begin{figure}\centering
 \includegraphics[width=8.5cm]{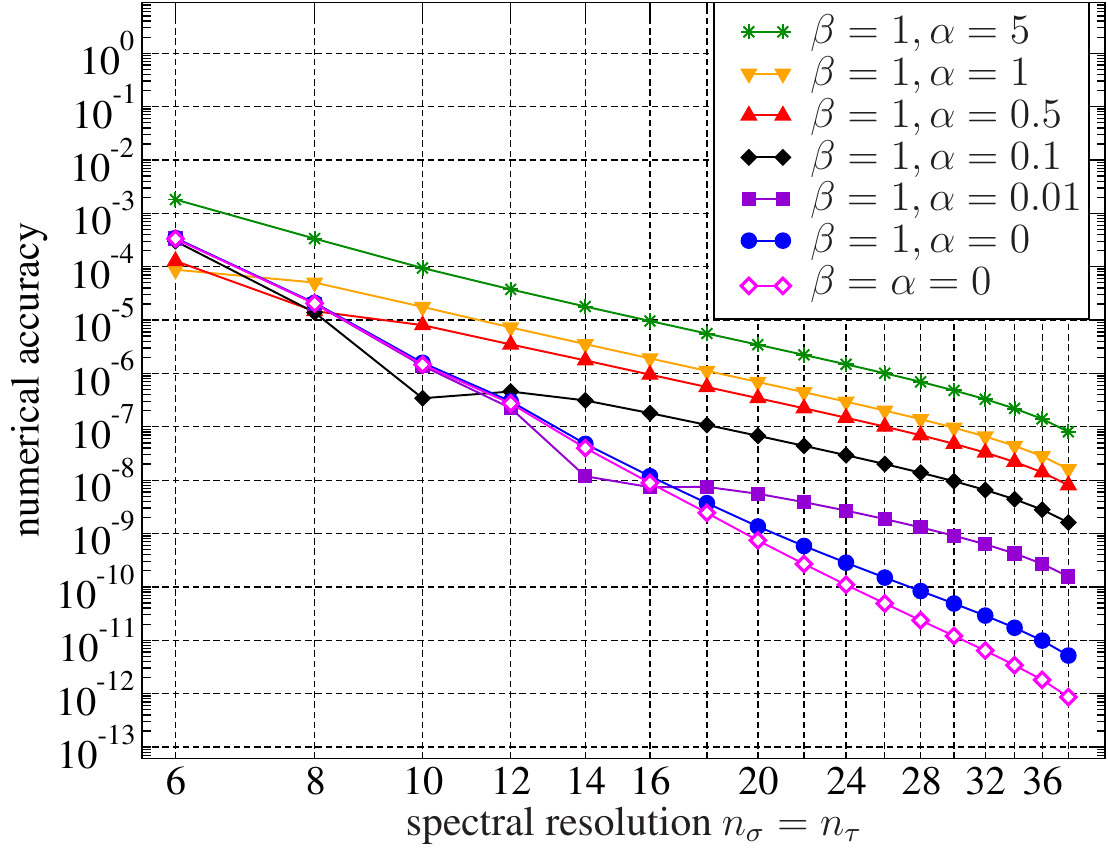}
 \caption{Convergence plot for the family of initial data \eqref{eq:iniseq} for several values of the parameters $\alpha$ and $\beta$. Note the double-logarithmic scale.\label{fig:alpha}}
\end{figure}

Numerical experiments show that the fully-pseudospectral scheme also converges in these cases. Note that, in order to determine the accuracy of the resulting solutions, we have to find a different measure than before, since no exact solution is available for these examples. To this end, we compare the function value $f(1,\rho_\mathrm{max})$ as obtained for some resolution with the corresponding value obtained for the highest resolution that we choose as $n_\sigma=n_\tau=42$. Since our numerical method effectively couples all gridpoints, any small error at one point usually spoils the accuracy at all other points. Hence the investigation of the behaviour at just one point is already a good measure for the overall accuracy.\footnote{Of course, we have to choose a point where the function value is not fixed by initial or boundary conditions. In our example, the values at $\tau=0$ are given in form of the initial values, while the values at $\rho=0$ are constant for our boundary condition $f_{,\tau}(\tau,0)=0$, so that $f$ is effectively known at $\rho=0$ as well. This explains our choice of the point $(\tau=1,\rho=\rho_\mathrm{max})$, which ensures that we look at function values as far away as possible from the region where $f$ is trivially known.}
It turns out that the curves in the resulting convergence plot are not straight lines as in the case of the above test solution, if we choose a logarithmic scale as in Fig.~\ref{fig:TestSol}. However, if we use a log-log plot instead, we again (roughly) obtain straight lines. This is shown in Fig.~\ref{fig:alpha}.
Based on this observation, we conclude that the error does not decay exponentially, but proportional to a (negative) power of the resolution, i.e.\ we have \emph{algebraic convergence} rather than the previous spectral convergence. This is exactly what one expects for pseudospectral methods if the solution is only $C^k$ with some finite $k$ rather than $C^\infty$ or analytic \cite{Boyd}. 

In the following, we have a closer look at the examples illustrated in Fig.~\ref{fig:alpha}. 
First we consider the case $\beta=1$ and choose several non-zero values for $\alpha$. The results can then be compared with the case $\alpha=0$. The corresponding curves in Fig.~\ref{fig:alpha} clearly show that the singularity in $\phi_2$ has a strong influence on the numerical accuracy. Without that singularity (i.e.\ for $\alpha=0$), the final error is below $10^{-11}$. However, even for the small value  $\alpha=0.01$, the final error is already about two orders of magnitude larger. If we further increase $\alpha$, then the singularity becomes stronger and the accuracy is further reduced. Finally, we consider the case $\alpha=\beta=0$, in which the next logarithmic term, namely the singularity in $\phi_3$, is removed as well. Compared to the simulations with $\alpha=0$, $\beta=1$, we gain about one further order of accuracy, and the error for the largest considered resolution is around $10^{-12}$. This improvement is smaller than the one achieved by going from $\alpha=0.01$ to $\alpha=0$ (in the case $\beta=1$). Hence we conclude that the singularity in $\phi_3$ is already of sufficiently high order to have only a relatively small influence on the result. Nevertheless,  the convergence curve corresponding to $\alpha=\beta=0$ is also a straight line in the log-log plot, which illustrates that the pseudospectral methods still clearly notice the fact that the solution has a limited regularity.

In summary, we see that the numerical accuracy very much depends on our choice of initial data. If we want highly accurate results, we can restrict ourselves to data satisfying the first two, or even the first three regularity conditions (rather than only the first), thus leaving logarithmic singularities only at high orders. Nevertheless, even for our most inaccurate example (obtained for $\alpha=5$, $\beta=1$), the final error of about $10^{-7}$ is still better than what some other numerical methods typically achieve. We could also just choose a larger spectral resolution, which would further reduce the error, but this would somehow be against the spirit of pseudospectral methods, where one wants to obtain very good results for a moderate number of gridpoints.

\begin{figure}\centering
 \includegraphics[width=8.5cm]{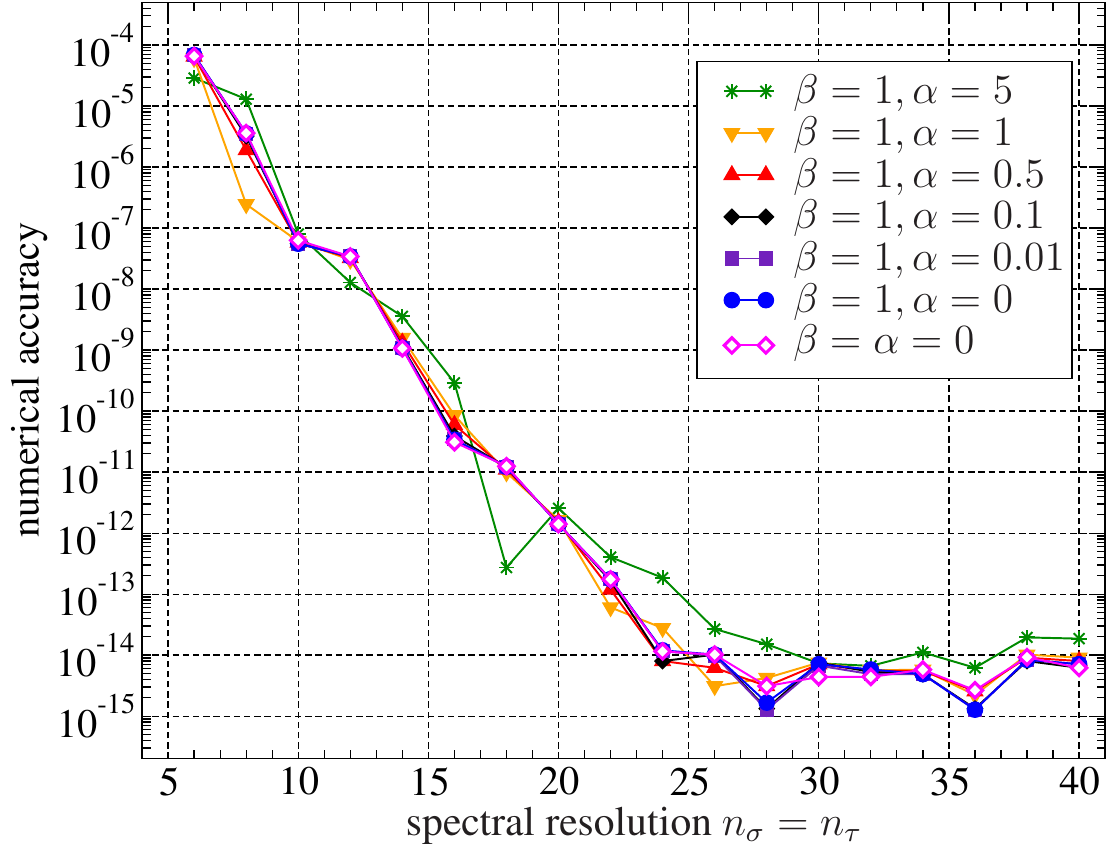}
 \caption{Convergence plots for the same examples as in Fig.~\ref{fig:alpha}, but this time on the smaller time interval $0\le\tau\le 0.9$.\label{fig:alphataumax}}
\end{figure}

Finally, we want to verify that the observed algebraic convergence for the solutions with the above initial data \eqref{eq:iniseq} is indeed caused by the logarithmic singularities at $\Scri^+$ and does not result from some other unpleasant feature of the particular solutions. For that purpose, we solve the same initial value problems from Fig.~\ref{fig:alpha} again, but this time on the smaller domain $0\le\rho\le\rho_\mathrm{max}$, $0\le\tau\le0.9$, which does not contain the upper boundary $\tau=1$. Since the solutions should be well-behaved in the entire new domain, we should find better convergence properties. This is what we indeed observe in Fig.~\ref{fig:alphataumax}, where the errors approximately follow straight lines in a logarithmic plot (rather than the previously used double-logarithmic scale). Moreover, the convergence curves for the different parameter choices overlap, and we find comparable errors for all examples. Saturation with final errors of about $10^{-14}$ close to machine-accuracy is reached with between $25$ and $30$ gridpoints in spatial and time directions.
This shows that we can restore all the nice features that pseudospectral methods usually have, if we exclude the logarithmic singularities from our computational domain.

\section{Including the event horizon}\label{sec:horizon}
\subsection{On three modifications of the compactified coordinates}

In the previous discussion our attention was focused on an accurate description of the cylinder at spacelike infinity and on solving the conformally invariant wave equation in a neighbourhood of this cylinder. However, we have already seen that our coordinates do not allow us to include the event horizon into our considerations, due to the coordinate singularity at $\rho=1$, which prevents access from the horizon at $r=1$ (which is formally located at $\rho=1/(1-\tau)>1$). This singularity is introduced with the coordinate transformation \eqref{eq:trans2}, since, for points close to the horizon, we have $r<1$ and $\rho>1$, such that the integration interval $r\le s\le\rho$ in \eqref{eq:trans2} then contains the point $s=1$ at which $1/F(s)$ is singular. 
In the following we will --- in several steps --- construct alternative coordinates which are suitable for solving the wave equation in a domain that contains both the cylinder and the event horizon.

\begin{figure}\centering
 \includegraphics[width=\textwidth]{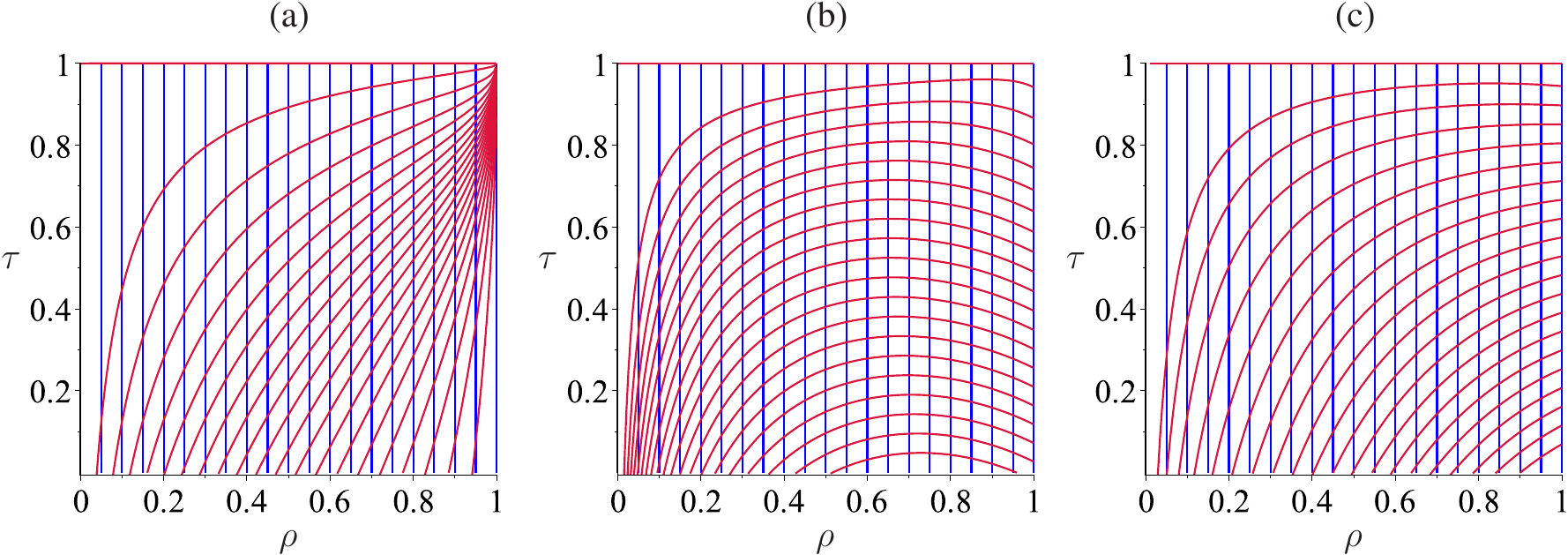}
 \caption{Radial null geodesics in the three $\rho$-$\tau$ coordinate systems defined via the coordinate transformations \eqref{eq:trans3} [panel (a)], \eqref{eq:trans4} [panel (b)] and \eqref{eq:trans5} [panel (c), shown for the parameter choice $d=3$].\label{fig:geodesics}}
\end{figure}

In a first step, we can try to modify the transformation equation $r=\rho(1-\tau)$. This part of the coordinate transformation achieves the blow up of $i^0$, and the important point is that both $\rho=0$ ($i^0$) and $\tau=1$ ($\Scri^+$) correspond to $r=0$, i.e.\ to an infinite value of the radial isotropic Schwarzschild coordinate $\tilde r=m/(2r)$, cf.~\eqref{eq:trans1}. A modified transformation equation should still have this property, but at the same time avoid that $\rho$-values greater than $1$ are required. This could be achieved by replacing \eqref{eq:trans2} with the following transformation, \begin{equation}\label{eq:trans3}
 t=\int_r^\rho\frac{\dd s}{F(s)},\quad r=\frac{\rho(1-\tau)}{1-\rho\tau},
  \quad\textrm{where}\quad
   F(s)=\frac{s^2(1-s)}{(1+s)^3}.
\end{equation}
With the new formula for $r$, the boundaries $\rho=0$ and $\tau=1$ still correspond to $r=0$, but $\rho=1$ does now imply $r=1$. Hence the resulting coordinates cover the horizon, which will be located at $\rho=1$. It is easy to check that the corresponding metric is regular for $\rho<1$, but not invertible at the horizon $\rho=1$. Moreover, radial null geodesics are irregular near the horizon: while one family of null geodesics is simply given by $\rho=\mathrm{constant}$, the geodesics in the other family all accumulate at the point $\rho=\tau=1$, see Fig.~\ref{fig:geodesics}(a). 
This is similar to the behaviour of null geodesics in advanced or retarded Eddington-Finkelstein coordinates, in which one class of geodesics is well-behaved, but the other class only approaches the horizon asymptotically. In our compactified picture, this asymptotic approach is mapped to the finite coordinate value $\tau=1$.

In order to improve our coordinates, we can choose a new time coordinate adapted to the behaviour of geodesics near the accumulation point. From the normalisation condition $g_{ab}\frac{\dd x^a}{\dd\lambda}\frac{\dd x^b}{\dd\lambda}=0$ for null geodesics, one can derive that the geodesics near the point $\rho=\tau=1$ are approximately given by $\tau\approx 1-c\sqrt{1-\rho}$, $c=\mathrm{constant}$. Based on this observation, we introduce a new time coordinate $\tau'$ via
\begin{equation}\label{eq:trans4a}
 \tau=1-(1-\tau')\sqrt{1-\rho}.
\end{equation}
If we only do this coordinate change, then geodesics in the resulting coordinate system would no longer be accumulated at a single point, but they would have an infinite slope at the horizon. This is caused by the term $\sqrt{1-\rho}$ in the coordinate transformation. We can solve this problem by also changing our $\rho$-coordinate via
\begin{equation}\label{eq:trans4b}
 \rho'=1-\sqrt{1-\rho},
\end{equation}
which is still defined in the interval $[0,1]$. If we combine our earlier transformation \eqref{eq:trans3} with the subsequent transformations \eqref{eq:trans4a}, \eqref{eq:trans4b}, then we can express the complete transformation as follows (where we skip the primes),
\begin{equation}\label{eq:trans4}
 t=\int_r^{\rho(2-\rho)}\frac{\dd s}{F(s)},\quad r=\frac{\rho(1-\tau)}{w+\rho(1-\tau)},
  \quad w:=\frac{1-\rho}{2-\rho}
\end{equation}
with the same function $F(s)$ as before. The resulting geodesics are now regular and reach the horizon at a finite slope, see Fig.~\ref{fig:geodesics}(b). 
However, these coordinates still have an unpleasant feature: the behaviour of the geodesics close to the initial surface $\tau=0$ shows that this surface is not everywhere spacelike. In fact, it turns out that $\tau=0$ is spacelike only for $\rho<0.735947731\dots$
In order to prescribe initial data for the wave equation, we therefore either have to choose a different, spacelike surface or find yet another coordinate system in which $\tau=0$ is spacelike.

We choose the latter approach and include an additional term in the coordinate transformation \eqref{eq:trans4a}, which we replace with
\begin{equation}
 \tau=1-(1-\tau')\sqrt{1-\rho}\left[1+d\left(1-\sqrt{1-\rho}\right)\right],
\end{equation}
where $d$ is a parameter. This allows us to control the shape of the $\tau'$-coordinate lines at one further order. If we again combine this with the previous transformations (and drop the primes), then we can write down the complete transformation as follows,
\begin{equation}\label{eq:trans5}
 t=\int_r^{\rho(2-\rho)}\frac{\dd s}{F(s)},\quad r=\frac{\rho(1-\tau)}{w+\rho(1-\tau)},
  \quad w:=\frac{1-\rho}{(2-\rho)(1+d\rho)},
\end{equation}
which differs from \eqref{eq:trans4} by an additional term in the definition of $w$.
Obviously, our new coordinates contain the previous ones as the special case $d=0$. However, if we restrict ourselves to parameter values $d>1/2$, then the initial surface $\tau=0$ is spacelike as required, see the example in Fig.~\ref{fig:geodesics}(c).

\begin{figure}\centering
 \includegraphics[width=7.3cm]{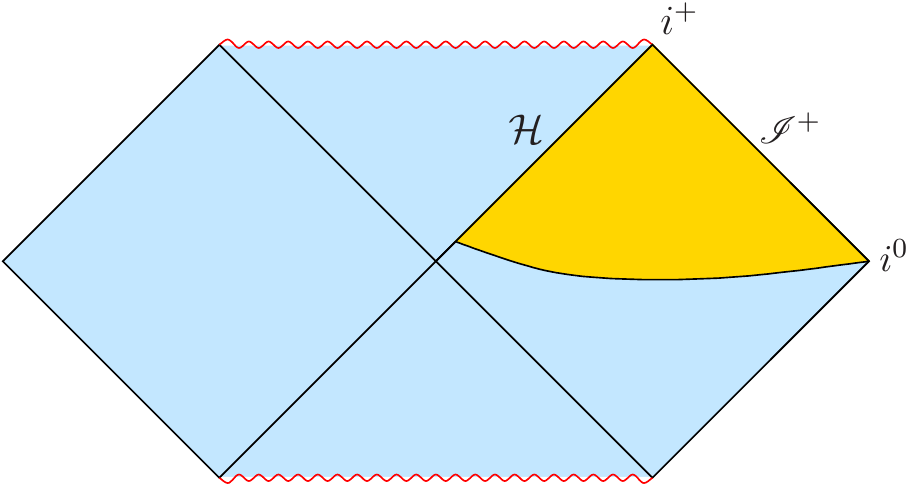}
 \caption{Sketch of the domain that is covered by the coordinates introduced in \eqref{eq:trans5}. Note that the exact form of the lower boundary depends on the parameter $d$. (The picture shows how this curve would approximately look like for $d=2$, provided that the Penrose diagram is constructed by compactifying the Kruskal null coordinates with $\mathrm{arctan}$-functions.)\label{fig:ConfDiag2}}
\end{figure}

If we consider the new coordinates in the domain $0\le\rho\le1$ and $0\le\tau\le 1$, then we describe a part of the Schwarzschild spacetime as shown in Fig.~\ref{fig:ConfDiag2}. 
Similarly to our earlier coordinates, the boundary $\rho=0$ corresponds to the cylinder $i^0$, and $\Scri^+$ is located at $\tau=1$. Now we also cover the event horizon $\mathcal H$, which is at the boundary $\rho=1$. Note that, in our earlier coordinates, the initial Cauchy surface $\tau=0$ corresponded to a vanishing isotropic Schwarzschild time, $\tilde t=0$, cf.\ \eqref{eq:trans2} or \eqref{eq:texplicit}. For the present coordinates, however, $\tau=0$ is a different hypersurface, parts of which can have positive values of $\tilde t$ and other parts negative. Therefore, the lower boundary of the relevant domain in Fig.~\ref{fig:ConfDiag2} is no longer a straight line (as in the corresponding picture in Fig.~\ref{fig:ConfDiag}), and we effectively solve initial value problems with data on a different hypersurface.

\subsection{The conformally invariant wave equation}

We express the Schwarzschild metric in terms of our (new) coordinates $\tau$ and $\rho$ as defined in \eqref{eq:trans5}. The result is the following conformal metric,
\begin{equation}\fl
 g = \frac{2 w}{\rho}B\,\dd\rho\,\dd\tau
     +\frac{1-\tau}{\rho^2}B\Big[(1-\tau)[w+\rho(1-\tau)]^2B-2W\Big]\dd\rho^2
     -\dd\sigma^2,
\end{equation}
where $w$ was defined in \eqref{eq:trans5}, and the quantities $B$ and $W$ are the following abbreviations,
\begin{equation}\fl
 B:=\frac{2(1-\rho)F(r)}{(1-\tau)^2F(\rho[2-\rho])}
   \equiv \frac{2(1-r)}{[w+\rho(1-\tau)]^2(2-\rho)^2(1-\rho)}
   \left(\frac{1+\rho(2-\rho)}{1+r}\right)^3,
\end{equation}
\begin{equation}
 W:=w-\rho\frac{\dd w}{\dd\rho}
  \equiv w+\frac{\rho\left[1+d+d(1-\rho)^2\right]}{(2-\rho)^2(1+d\rho)^2}.
\end{equation}
The conformal factor is the same as before, given in \eqref{eq:Theta}.

Using this new form of the conformal metric, we can again formulate the conformally invariant wave equation. The result is
\begin{eqnarray}
 (1-\tau)\Big[2W-(1-\tau)[w+\rho(1-\tau)]^2B\Big]f_{,\tau\tau}
 +2\rho wf_{,\tau\rho}\nonumber\\
  +2\left[(1-\tau)[w+\rho(1-\tau)]B\left(w \frac{1-2r}{1-r^2}+\rho(1-\tau)\right)-W\right]f_{,\tau}\nonumber\\
  +\frac{4 r w^2B}{(1+r)^2} f=0.
\end{eqnarray}

Similarly to our previous considerations, we can again derive boundary conditions for regular solutions from the wave equation. In the limit $\rho\to0$, the wave equation again becomes an intrinsic equation whose solutions blow up for $\tau\to1$, unless we restrict ourselves to solutions that are constant on $i^0$, which is guaranteed if we impose the same regularity condition on our initial data as before, namely,
\begin{equation}\label{eq:newreg}
 f_{,\tau}(0,0)=0.
\end{equation}
This is what we will do in the following. Similarly, one could repeat the entire earlier discussion of logarithmic singularities at higher orders and impose further conditions to avoid these. However, here it is better to choose a numerical domain that does not contain $\Scri^+$. The reason is that, in the current setting with a domain that includes the event horizon $\mathcal H$, we do not only have to cope with logarithmic singularities at $I^+$ ($\rho=0$, $\tau=0$), but also with a physical singularity at future timelike infinity $i^+$ ($\rho=\tau=1$). (Recall that the curvature singularity of the Schwarzschild solution, indicated by the upper wiggly line in Fig.~\ref{fig:ConfDiag2}, touches the point $i^+$ in the compactified picture.) Hence there is no hope that solutions to the wave equation will be regular at $i^+$, if the underlying spacetime is not. Consequently, we will only solve the problem in a smaller domain that excludes this singularity. To this end, we restrict the coordinates to a rectangular domain of the form
\begin{equation}
 0\le\rho\le 1,\quad 0\le\tau\le\tau_\mathrm{max}<1.
\end{equation}
With this choice, we obtain the solution in a region that contains both the cylinder $i^0$ (at $\rho=0$) and the horizon $\mathcal H$ (at $\rho=0$), but not $\Scri^+$ (at $\tau=1$). As a consequence, the logarithmic singularities at $I^+$ are not relevant for our discussion as this point is excluded from our domain too. (As we have seen before, the logarithmic singularities are too mild to cause problems in domains away from $I^+$, cf.~Fig.~\ref{fig:alphataumax}.) The singularity at $i^+$, on the other hand, could be strong enough to influence the numerical accuracy if $\tau_\mathrm{max}$ is too close to $1$. Hence we will need to investigate what values of $\tau_\mathrm{max}$ lead to acceptable results.

Besides $i^0$, the wave equation also becomes an intrinsic equation on the horizon. This equation can be integrated to show that the following linear combination of first-order derivatives is constant on the horizon,
\begin{equation}
 \mathcal H: \quad
 [1-2d(1-\tau)]f_{,\tau}-2(1+d)f_{,\rho}=\mathrm{constant}.
\end{equation}
For our numerical purposes, however, this is not important, as we just impose the wave equation itself at $\rho=1$ (which is a characteristic boundary, so that no additional conditions are required there).

\subsection{Numerical studies}\label{sec:numstud}

\begin{figure}\centering
 \includegraphics[width=8.5cm]{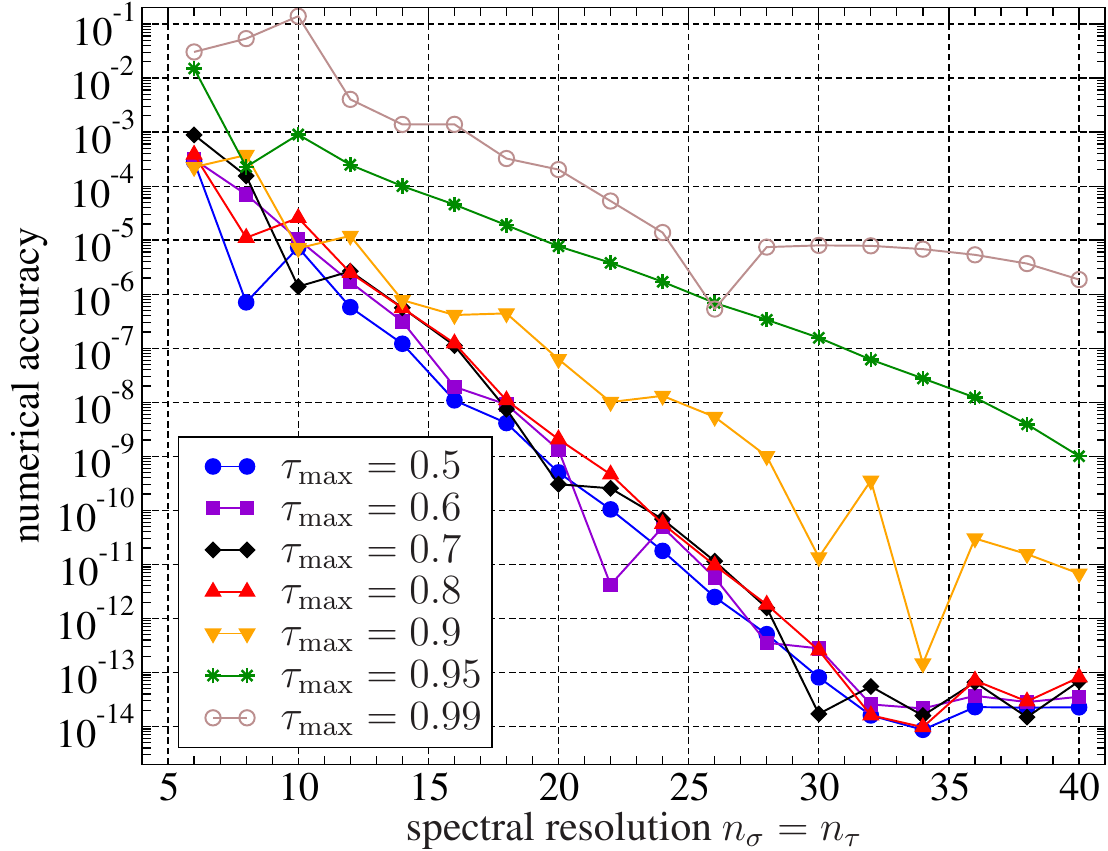}
 \caption{Convergence plot for the initial data \eqref{eq:inidat}. The parameter $d$ in the coordinates has been chosen as $d=2$.\label{fig:d-metric}}
\end{figure}

We intend to test how accurately the conformally invariant wave equation can be solved with the fully pseudospectral method. Note that in our new coordinates, we cannot try to reconstruct the test solution \eqref{eq:testsol}, as this diverges at the horizon, i.e.\ for $r\to 1$, and for which even initial data at $\tau=0$ would be irregular. Instead we have to choose some regular initial data. As an example, consider the following data,
\begin{equation}\label{eq:inidat}
 \tau=0:\quad f=\cos(2\rho),\quad f_{,\tau}=\sin(\rho),
\end{equation}
which satisfy the regularity condition \eqref{eq:newreg}.
In order to test how close we can come to the singularity at $\tau=1$, we solve the corresponding initial value problem on different domains, obtained by choosing different values for $\tau_\mathrm{max}$. To measure the numerical accuracy, we  compare the function values $f(\tau_{\max},1)$ as obtained for various spectral resolutions to the value obtained for the highest resolution of $42\times 42$. The resulting convergence plot is shown in Fig.~\ref{fig:d-metric}.

For $\tau_\mathrm{max}=0.5,0.6,0.7,0.8$, we observe spectral convergence and highly accurate solutions with error saturation below $10^{-13}$. The errors in all these examples are comparable, i.e.\ the particular value of $\tau_\mathrm{max}$ has no big influence on the accuracy as long as the numerical domain is sufficiently far away from $i^+$. For $\tau_\mathrm{max}=0.9$, however, there is a considerable loss of accuracy, compared to the previous examples: the error is about two orders of magnitude larger. Nevertheless, the smallest error as obtained for the highest resolution is below $10^{-11}$, so that we still have very accurate results.
With further increasing $\tau_\mathrm{max}$ to $0.95$ we loose two further orders of magnitude, and for $\tau_\mathrm{max}=0.99$ the final error is as large as $2\times 10^{-6}$. This shows that, as expected, our numerical scheme depends quite sensitively on the proximity to the singularity. However, unless we are really close to $\tau=1$, very accurate solutions are found.

\section{Discussion}\label{sec:discussion}

We have studied the conformally invariant wave equation on a Schwarzschild spacetime with particular consideration of an appropriate treatment of spacelike infinity $i^0$ as a cylinder $I$. This problem can be regarded as a toy model for the general behaviour of more complicated equations (e.g., the spin-2 system or the full Einstein equations) and their solutions near the cylinder.  In particular, our present studies extend the previous discussions of the Minkowski wave equation in \cite{FrauendienerHennig2014} to a nontrivial background metric. 

Our approach should be compared to the ``standard'' way of solving problems on compactified spacetimes, where data are given on a hyperboloidal or characteristic initial slice. The data are then chosen somewhat arbitrarily such that everything extends smoothly to future null infinity. However, in this way, the connection to spacelike infinity is completely lost, and there is no way to say anything about past null infinity.

We found that we can numerically solve the conformally invariant wave equation for the Schwarzschild spacetime with the fully pseudospectral scheme that we also used in the Minkowski case. Moreover, for sufficiently regular solutions, we observe the same numerical properties: we achieve highly accurate solutions close to machine accuracy with a moderate number of gridpoints, and the error decreases exponentially with the spectral resolution, i.e.\ we have spectral convergence. However, as opposed to the Minkowski case, the solutions are generally not regular if future null infinity $\Scri^+$ is part of the numerical domain. Indeed, in the present Schwarzschild case, the solutions generally develop logarithmic singularities at $I^+$, 
where $I$ and $\Scri^+$ approach each other.
As a consequence, the numerical convergence is only algebraic in those cases. Nevertheless, the numerical approximations turn out to be still very accurate. 
The analysis of the behaviour of solutions along the cylinder shows that logarithmic singularities can appear at $I^+$ in derivatives of every order. These singularities can be eliminated order by order as discussed in the appendix which has the consequence that the solution can be made $C^k$ at $I^+$ for arbitrary natural number $k$ by imposing sufficiently many conditions at $I^0$. The interesting feature of the fully pseudospectral method is that the finite differentiability is directly reflected in the convergence behaviour of the solution. In this way, we have a direct way of \emph{controlling the smoothness on null infinity} by imposing conditions at $I^0$ \emph{on the initial data}. In this sense
the influence of logarithmic singularities on the convergence properties of our numerical method is actually something very positive. In this way, the numerical simulations can be used to analyse the degree of regularity of the solution for any given set of initial data. No other numerical method may be able to provide this useful information.

Overall, we see that the fully pseudospectral scheme is very well-suited for treating the conformally invariant wave equation on a Schwarzschild background, even in the presence of logarithmic singularities. Hence we have found yet another application for which this method works very well, and we can expect that more complicated equations and more general problems can be successfully studied with the same approach in the future.

Another important object of further analysis would be to generalise the present considerations to numerical domains that include (a part of) past null infinity. As discussed in the construction of coordinates in Sec.~\ref{sec:coords}, our $\rho$-$\tau$ coordinates only cover the ``future half'' of the Schwarzschild spacetime. Moreover, coordinates based on the Kruskal extension introduce singularities at the cylinder. Hence the currently available coordinate systems are either irregular or do not cover past null infinity, and it would first be necessary to identify new types of well-behaved coordinates.

\appendix

\section{A general recursion formula for the expansion at spacelike infinity}\label{sec:recursion}

In Sec.~\ref{sec:expansion} we have explicitly derived how the solution $f(\tau,\rho)$ behaves near spacelike infinity in the first four orders in $\rho$, described by the functions $\phi_0(\tau),\dots,\phi_3(\tau)$. Here we give a recursion for $\phi_n(\tau)$, which allows to study properties of the expansion in greater generality.

In order to obtain the desired formula, we multiply the wave equation \eqref{eq:CWE1} with $(1-\rho)[1+\rho(1-\tau)]^5$, which leads to an equation in which each derivative of $f$ is multiplied by a polynomial in $\rho$ (with $\tau$-dependent coefficients). Plugging the expansion \eqref{eq:expan} into this equation and comparing coefficients of different $\rho$-powers, we obtain a recursion formula. A certain complication is introduced by the fact that the coefficient polynomials in the reformulated wave equation are of up to seventh degree, which mixes quite a number of different functions $\phi_n$. Nevertheless, the somewhat involved formula that one obtains has a simple structure,
\begin{equation}\label{eq:oderecursion}
 (1-\tau^2)\ddot\phi_n+2(n-\tau)\dot\phi_n=R_n,
\end{equation}
where $R_n$ depends on the six previous $\phi$-functions $\phi_{n-1},\dots,\phi_{n-6}$. Hence, even just to start the iteration, we need more $\phi$-functions than we have computed above. 

For completeness, we give the full expression for $R_n$, which reads
\begin{eqnarray*}
 \fl
 R_n &=& 
 \Big((1- \tau) ( \tau^2+5  \tau-4) \ddot\phi_{n-1}+2 [ \tau^2+(5 n-9)  \tau+6-4 n] \dot\phi_n-4 (1- \tau) \phi_{n-1}\Big)\\
 \fl &&
 -(1- \tau) \Big((1- \tau)(\tau^2-19  \tau+5) \ddot\phi_{n-2}
 -2 [2  \tau^2+(10 n-37)  \tau+17-5 n] \dot\phi_{n-2}\\
\fl &&
 +4 ( \tau+2) \phi_{n-2}\Big)
 -(1- \tau)\Big(- \tau (1- \tau) ( \tau^2-26  \tau+26) \ddot\phi_{n-3}\\
\fl &&
 +2 [(10 n-46)  \tau^2+(55-10 n)  \tau-8] \dot\phi_{n-3}+12  \tau \phi_{n-3}\Big)\\
\fl &&
 -(1- \tau) \Big((1- \tau)^2(13  \tau^2-9  \tau-5)\ddot\phi_{n-4}
 +2 (1- \tau)[5(n-5)  \tau^2+6  \tau+18-5 n] \dot\phi_{n-4}\\
\fl &&
 +4 (3 \tau-2) \phi_{n-4}\Big)
 -(1- \tau)^2 \Big((1- \tau)^2(2\tau^2+3  \tau-4)\ddot\phi_{n-5}\\
\fl &&
 +2(1- \tau)[(n-6)\tau^2+(3 n-18)  \tau+22-4 n] \dot\phi_{n-5}-4\phi_{n-5}\Big)\\
\fl &&
 -(1- \tau)^5\Big((2\tau-1) \ddot\phi_{n-6}-2(n-7)\dot\phi_{n-6}\Big).
\end{eqnarray*}
The second-order equation \eqref{eq:oderecursion} for $\phi_n$ can, in principle, be solved with two integrations by rewriting it as
\begin{equation}\label{eq:recursion2}
 \frac{\dd}{\dd\tau}\left(\frac{(1+\tau)^{n+1}}{(1-\tau)^{n-1}}\dot\phi_{n}\right)
 =\left(\frac{1+\tau}{1-\tau}\right)^nR_n.
\end{equation}
Obviously, two new integration constants will be introduced at each order $n$.

If we assume that an appropriate choice of the integration constants (corresponding to a suitable subset of initial data) has eliminated all singularities up to the order $n-1$, then $R_n$ will be a regular function of $\tau$. If we further assume that this regular function has an expansion
\begin{equation}
R_n(\tau)=c_0+c_1(1-\tau)+c_2(1-\tau)^2+\dots, 
\end{equation}
near $\tau=1$, then the differential equation can be explicitly solved for $\phi_n(\tau)$. We observe that the contribution of the term $c_k(1-\tau)^k$ is regular as long as $k>n-1$. For
$k=0,1,\dots,n-1$, however, the results contain terms that behave like $(1-\tau)^n\ln(1-\tau)$ near $\tau=1$. Consequently, we expect weaker and weaker logarithmic singularities at infinitely many orders. 

This does not explicitly show how we have to restrict the initial data in order to eliminate the logarithmic singularities up to a given order, since that would require an explicit knowledge of the solution including the relevant integration constants. But it illustrates the structure of the singular terms, and it provides a systematic approach to computing $\phi$-functions up to arbitrary orders.

Finally, it is interesting to see how the singular behaviour results from the non-vanishing mass $m$ of the background Schwarzschild spacetime. Since the mass is absorbed into the definition of the coordinates $\tau$ and $\rho$, this is not directly possible with the previous equations. However, we can re-derive the wave equation in coordinates that leave $m$ in the metric and equations. To this end, we replace \eqref{eq:trans1} with $r=1/(2\tilde r)$, $t=2\tilde t$ and change the definition of $F(s)$ from \eqref{eq:trans2} to $F(s)=s^2(1-ms)/(1+ms)^3$. We can then perform a similar analysis of the resulting wave equation. In particular, it turns out that the differential equation \eqref{eq:oderecursion} needs to be replaced with
\begin{equation}\label{eq:oderecursion1}
 (1-\tau^2)\ddot\phi_n+2(n-\tau)\dot\phi_n=m\tilde R_n,
\end{equation}
where $\tilde R_n$ is a slightly modified version of the previous function $R_n$. Most importantly, we observe that the equation has a factor $m$ on the right-hand side. As long as $m\neq0$, the above analysis applies, which reveals singularities at infinitely many orders. However, for $m=0$, the right-hand side vanishes so that the equations for the different $\phi$-functions decouple. We can even solve them explicitly to obtain that
\begin{equation}
 \phi_n(\tau)=\left\{\begin{array}{ll}\displaystyle
               c_0+d_0\ln\left(\frac{1-\tau}{1+\tau}\right), & n=0\\[2.5ex]
               \displaystyle
               c_n+d_n\left(\frac{1-\tau}{1+\tau}\right)^n, &n>0
              \end{array}\right.
\end{equation}
with integration constants $c_n$, $d_n$. Evidently, the corresponding expansion has a singularity only at the lowest order --- exactly what we know about the Minkowski situation --- while singular terms at all other orders disappear.

\section*{Acknowledgments}
 We would like to thank Chris Stevens for commenting on the manuscript, and JF would like to thank the Department of Mathematics at the University of Oslo for hospitality. Part of this research was supported by the European Research Council through the FP7-IDEAS-ERC Starting Grant scheme, project 278011 STUCCOFIELDS. Moreover, we thank the anonymous referees for their valuable constructive comments.
 


\end{document}